\pdfoutput=1
\documentclass[aps,prd,amsmath,floats,floatfix, twocolumn,
superscriptaddress,nofootinbib,showpacs,longbibliography]{revtex4-2}

\usepackage[T1]{fontenc}
\usepackage[utf8]{inputenc}
\usepackage{lmodern}
\usepackage{mathrsfs}
\usepackage{amsfonts}

\usepackage{verbatim}

\usepackage{ragged2e}
\usepackage{subcaption} 
\usepackage{etoolbox}

\makeatletter
\pretocmd{\caption}{\justifying\captionsetup{singlelinecheck=false}}{}{}
\makeatother

\usepackage[dvipsnames, usenames]{xcolor}
\definecolor{linkcolor}{rgb}{0.0,0.3,0.5}
\usepackage[hypertexnames=false, unicode, colorlinks=true, linkcolor=linkcolor,
citecolor=linkcolor, filecolor=linkcolor,urlcolor=linkcolor,
pdfusetitle]{hyperref}

\usepackage[all]{hypcap}
\usepackage{graphicx}
\usepackage{xspace}
\usepackage{hyperref}
\usepackage{amssymb}

\usepackage[normalem]{ulem} 
\usepackage{bm} 

\usepackage{microtype}

\usepackage{blindtext}

\usepackage{orcidlink}

\graphicspath{%
  {figs/}%
}

\DeclareMathAlphabet{\mathpzc}{OT1}{pzc}{m}{it}

\newcommand{\pd}{\partial}

\begin{document}

\title{Gauge Boundary conditions to mitigate center-of-mass drift in BBH simulations}

\newcommand{\Cornell}{\affiliation{Cornell Center for Astrophysics
    and Planetary Science, Cornell University, Ithaca, New York 14853, USA}}
\newcommand\CornellPhys{\affiliation{Department of Physics, Cornell
    University, Ithaca, New York 14853, USA}}
\newcommand\Caltech{\affiliation{Theoretical Astrophysics 350-17, California Institute of
    Technology, 1200 E California Boulevard, Pasadena, CA 91125, USA}}
\newcommand{\AEI}{\affiliation{Max Planck Institute for Gravitational Physics
(Albert Einstein Institute), D-14476 Potsdam, Germany}}
\newcommand{\UMassD}{\affiliation{Department of Mathematics,
    Center for Scientific Computing and Visualization Research,
    University of Massachusetts, Dartmouth, MA 02747, USA}}
\newcommand\UMiss{\affiliation{Department of Physics and Astronomy,
    University of Mississippi, University, MS 38677, USA}}
\newcommand{\Bham}{\affiliation{School of Physics and Astronomy and Institute
    for Gravitational Wave Astronomy, University of Birmingham, Birmingham, B15
    2TT, UK}}
\newcommand{\Perimeter}{\affiliation{Perimeter Institute for Theoretical Physics, Waterloo, ON N2L2Y5, Canada}}

\author{Dongze Sun
\orcidlink{0000-0003-0167-4392}}
\email{dzsun@caltech.edu}
\Caltech
\author{Sizheng Ma~\orcidlink{0000-0002-4645-453X}}
\Perimeter
\author{\\ Mark A.\ Scheel~\orcidlink{0000-0001-6656-9134}}
\Caltech
\author{Saul A. Teukolsky
\orcidlink{0000-0001-9765-4526}}
\Caltech
\Cornell

\hypersetup{pdfauthor={Sun et al.}}

\date{\today}

\begin{abstract}
Long-term numerical relativity (NR) simulations of binary black hole (BBH) systems in the Spectral Einstein Code (SpEC) code exhibit an unexpected exponential drift of the center-of-mass (CoM) away from the simulation's origin.
In our work, we analyze this phenomenon and demonstrate that it is not a physical effect but rather a manifestation of a gauge artifact.
The origin of this drift is the reflection of the gauge waves off the outer boundary of the computational domain.
These reflections are introduced by inaccuracies in the gauge boundary condition, specifically, the application of the Sommerfeld condition to the time derivative of the gauge fields. 
Such an approach fails to completely suppress or correctly absorb the outgoing modes, thereby generating artificial feedback into the simulation.
To mitigate this problem, we introduce a modified boundary condition that incorporates an explicit CoM correction source term designed to counteract the CoM motion.
Our numerical experiments, performed with the SpEC code, reveal that this new boundary treatment reduces the CoM drift by several orders of magnitude compared to the standard implementation, and does not introduce any unwanted physical artifacts.
\end{abstract}

\maketitle

\section{Introduction} \label{sec:intro}
Numerical relativity (NR) provides the most accurate solution for the dynamics and gravitational-wave emission of binary black holes (BBHs).  
The first stable BBH evolutions were obtained by Pretorius \cite{Pretorius:2005gq} in the generalized-harmonic formulation, followed soon after by the moving-puncture breakthroughs of Campanelli et al. \cite{Campanelli:2005dd} and Baker et al. \cite{Baker:2005vv}.
These NR results play a crucial role in understanding the dynamics of these astrophysical phenomena, and the behavior of gravity in extreme regimes
\cite{TheLIGOScientific:2016src,LIGOScientific:2019fpa,LIGOScientific:2021sio}.

\begin{figure}[thb]
\centering
\includegraphics[width=0.45\textwidth]{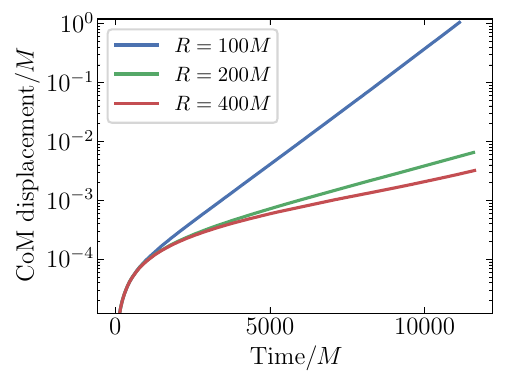}
\caption{\justifying The CoM displacement from the origin for different outer boundary radii. The system is equal mass and non-spinning. }
\label{fig:ExpFit}
\end{figure}

Despite major successes, long-duration NR simulations remain sensitive to how the artificial outer boundary is treated. 
In generalized-harmonic (GH) evolutions with the Spectral Einstein Code (SpEC) \cite{SpECwebsite}, an
exponential drift of the BBH system's coordinate center-of-mass (CoM) has been consistently observed \cite{Szil_gyi_2015}.
Figure~\ref{fig:ExpFit} illustrates this drift, which is a steadily increasing displacement of the CoM from the coordinate origin over time.
It is unclear whether this boundary issue would occur for moving-puncture/BSSN codes; unlike SpEC, these codes generally place the boundary very far out so that it is causally disconnected over the simulation time \cite{Pollney_2009,Hinder_2010,Etienne_2014,Mewes_2018}.

To uncover the cause of this spurious CoM
behavior, Ref.~\cite{Szil_gyi_2015}
manipulated the outer boundary radius and observed that the
exponential growth rate, denoted by $\sigma$, decreases monotonically
with increasing boundary distance as $\sigma \propto 1/R^{1.45}$.
This sensitivity to the boundary location strongly indicates that
the drift is related to a coupling between the evolving fields and
the computational boundary. This effect makes it infeasible to
conduct long-term NR simulations with small outer boundary radii.
To prevent the drift from affecting the simulation and waveform
extraction, the outer boundary is typically set  at radii on the
order of $10^3\,M$ \cite{Boyle_2019}, where $M$ is the total mass
of the system.  Although such large boundaries mitigate the drift,
they also come with an increase in computational expense.
Therefore, it is imperative to thoroughly investigate and address
this issue to enable more efficient long-term NR simulations.

In our study, we probe this effect and find that the CoM drift is
predominantly a gauge effect arising from the reflection of gauge
waves off the outer boundary.  These reflections are caused by the
imperfect non-reflective Sommerfeld boundary condition applied to
the time derivative of the gauge variables.  Essentially, while
the Sommerfeld condition is non-reflective for $l=0$ outgoing
radiation modes, any nonzero displacement of the CoM will generate
higher-order outgoing radiation modes that can be reflected by
the Sommerfeld condition.  These reflected waves caused by the
Sommerfeld condition then propel the CoM
with exponential acceleration.  Since the Sommerfeld condition is
only imposed on the gauge degrees of freedom of the fields, the physical CoM defined from global Poincaré charges \cite{TDray_1984,TDray_1985,Keefe:2021} is not affected and does not exhibit exponential drift, as we will
show in Sec.~\ref{sec:gauge}.

Motivated by these insights, we propose an improved boundary
treatment that mitigates the CoM drift.
We add an additional CoM correction source term to the boundary condition.
This term is formulated to effectively dampen the CoM velocity or realign the CoM with the coordinate origin.
Numerical tests using our modified boundary condition demonstrate a significant reduction in the drift amplitude, by several orders of magnitude relative to the original setup.
Thus, our modification not only clarifies the origin of the CoM drift but also provides a practical solution for performing long-term BBH simulations with smaller outer boundary radii and reduced computational cost.

In this paper, we use lower-case Latin indices from the beginning of the alphabet to denote four-dimensional spacetime quantities, whereas lower-case Latin indices from the middle of the alphabet are spatial.
The coordinate CoM is defined using the Newtonian formula as
\begin{equation}
\bm{\mathrm{r_s}}=\frac{m_1 \bm{\mathrm{r_1}}+m_2 \bm{\mathrm{r_2}}}{m_1+m_2},
\end{equation}
where $\bm{\mathrm{r_1}}$ and $\bm{\mathrm{r_2}}$ are the coordinates of the center of the two black holes, which are defined as the surface-area weighted average of the location of the apparent horizons \cite{Boyle_2019}.

The remainder of this paper is organized as follows.
In Sec.~\ref{sec:gauge}, we show that the CoM drift is a gauge effect by inspecting the physical charges.
In Sec.~\ref{sec:Sommerfeld}, we review the Sommerfeld gauge boundary condition that is used in SpEC, and we discuss why this boundary condition leads to the exponential drift of the CoM.
In Sec.~\ref{sec:EoM}, we derive the equation of motion of the CoM under given gauge boundary conditions.
Following this, we present the new boundary conditions that significantly damp the CoM drift in Sec.~\ref{sec:NewBC}.
And we show that the gauge boundary conditions do not affect the resulting gauge-fixed waveform in Sec.~\ref{sec:waveform}.
In Sec.~\ref{sec:alternative}, we discuss the potential of using higher-order boundary conditions, gauge constraint damping boundary condition, NR-post-Minkowskian matching, or Cauchy-Characteristic matching (CCM) as potential alternative solutions to address the CoM drift.
In Sec.~\ref{sec:conclusions}, we provide a few closing remarks.
The BBH simulations used in this paper are performed by the SpEC code \cite{SpECwebsite}.

\begin{figure}[thb]
\centering
\includegraphics[width=0.48\textwidth]{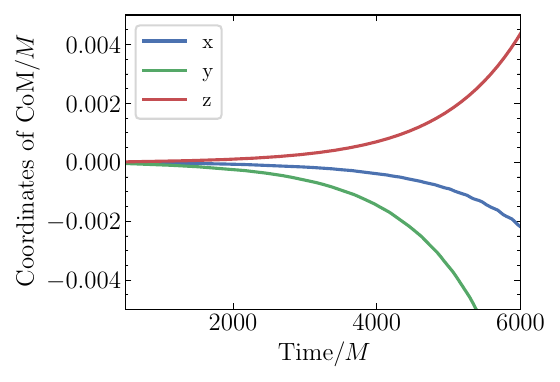}
\includegraphics[width=0.48\textwidth]{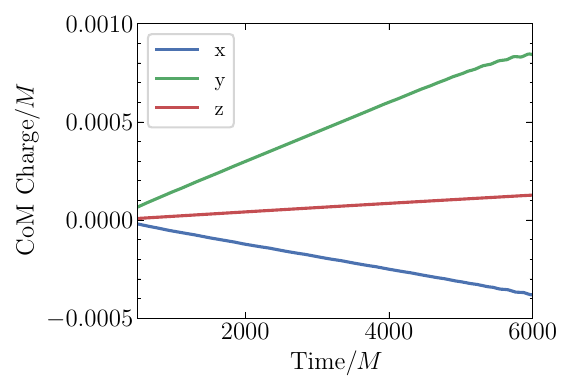}
\caption{\justifying The upper panel shows the evolution of CoM coordinates
and the lower panel shows their corresponding CoM charge. The outer
boundary radius is set as $100\,M$.}
\label{fig:charge}
\end{figure}

\section{CoM drift as a gauge effect}\label{sec:gauge}
To identify whether the CoM drift is a physical or gauge effect, we inspect the Poincar\'e charges that are associated with the global conserved physical quantities. These charges are gauge‑invariant up to the well‑understood Bondi-van~der~Burg-Metzner-Sachs (BMS) group freedom \cite{Sachs:1962wk,Sachs:1962zza,Bondi:1962px}. We will address the BMS frame ambiguity in Sec.~\ref{sec:waveform}.
The Poincar\'e charges are given by the integral of the Bondi mass aspect $m_\text{B}$, the Lorentz aspect $N$, and the energy moment aspect $E$ over the celestial two-sphere (future null infinity $\mathscr{I}^+$ from the source) \cite{Bondi:1962px}.
These functions are given in the Moreschi-Boyle convention \cite{Boyle:2015nqa,Keefe:2021,Keefe:2022} using the Bondi-Sachs coordinates $\{u,r,\theta^A\}$ as
\begin{equation}
\begin{aligned}
m_\text{B}(u, \theta^A) &\equiv -\mathrm{Re}\left[\Psi_2 + \sigma \dot{\bar{\sigma}} \right], \\
N(u, \theta^A) &\equiv -\left( \Psi_1 + \sigma \eth \bar{\sigma} + u \eth m + \frac{1}{2} \eth (\sigma \bar{\sigma}) \right), \\
E(u, \theta^A) &\equiv N + u \eth m \\
&= -\left( \Psi_1 + \sigma \eth \bar{\sigma} + \frac{1}{2} \eth (\sigma \bar{\sigma}) \right).
\end{aligned}
\end{equation}
Here $\Psi_i$ are the Weyl scalars, $\sigma$ is the Newman-Penrose shear, and an overbar represents complex conjugation.
For a quantity $\eta$ of spin weight $s$, the spin-weight raising operator $\eth$ can be written as \cite{Newman:1966ub, Boyle:2016tjj}
\begin{equation}
\eth\eta = -\frac{1}{\sqrt{2}}(\sin\theta)^s\left(\frac{\partial}{\partial\theta}+\frac{i}{\sin\theta}\frac{\partial}{\partial\phi}\right)\left((\sin\theta)^{-s}\eta\right).
\end{equation}

With these functions defined at $\mathscr{I}^+$, we then have the energy–momentum charge and complex dipole moment defined as
\begin{equation}
\begin{aligned}
P^a &= \frac{1}{4\pi}\int_{\mathscr{I}^+}n^a m_\text{B}d\Omega, \\
D^a+iJ^a&=\frac{1}{4\pi}\int_{\mathscr{I}^+}\bar{\eth}n^a Ed\Omega,
\end{aligned}
\end{equation}
where $n^a=\left(1,\sin\theta\cos\phi,\sin\theta\cos\phi,\cos\theta\right)$.
Here the real part $D$ defines the gauge-independent mass dipole, and the imaginary part $J$ defines the angular momentum.

A gauge-independent measurement of the CoM displacement, i.e., the CoM charge, can be defined as the mass dipole $D$ scaled by the total energy $P^0$ \cite{TDray_1984,TDray_1985,Keefe:2021}:
\begin{equation}
G^i\equiv\frac{D^i}{P^0}.
\end{equation}

To carry out this check, we use Cauchy-characteristic evolution (CCE) to extract the Weyl scalars and $\sigma$ at $\mathscr{I}^+$ \cite{Moxon_2020,moxon2021spectre}. 
The Cauchy evolution parts are performed using the SpEC code \cite{SpECwebsite}, and the CCE computations are performed using the SpECTRE Code \cite{spectrecode}.
CCE uses a worldtube from the Cauchy evolution of the Einstein field equations as the inner boundary to a characteristic evolution on a null foliation. 
The gravitational information is evolved to future null infinity, so one can extract Weyl scalars there unambiguously from any Cauchy evolution.

Fig.~\ref{fig:charge} shows the evolution of CoM coordinates alongside the corresponding CoM charges of an equal-mass and non-spinning system. 
The CoM charges are computed using the \texttt{scri} package \cite{scri}. 
The CoM charges in each direction shown in the lower panel are considerably smaller than the corresponding CoM coordinates shown in the upper panel, and they do not exhibit exponential growth. 
This is strong evidence that the exponential CoM drift primarily arises as a gauge effect caused by inaccuracies in the gauge boundary conditions, rather than a genuine physical motion.

\section{Sommerfeld boundary condition} \label{sec:Sommerfeld}
In this section, we first review the gauge boundary condition used in the SpEC code, and then use a propagating scalar wave as a toy model to show how reflections can be generated by this boundary condition.

\subsection{Gauge boundary condition in SpEC}
The gauge boundary condition used in SpEC is the Sommerfeld non-reflective boundary condition \cite{Sommerfeld,Rinne:2007ui}.
The Sommerfeld boundary condition acts on a general propagating tensor wave $\Psi$ as
\begin{equation}\label{eq:Sommerfeld}
L\Psi\doteq0,
\end{equation}
where
\begin{equation}\label{eq:SommerfeldOp}
L=\partial_t+\partial_r+\frac{1}{r},
\end{equation}
and $\doteq$ denotes equality at the boundary.

For gravitational waves, the spacetime metric can be approximately decomposed as a background part $g_{ab}^\text{(B)}$ that changes “adiabatically”, and a propagating wave part $h_{ab}$: 
\begin{equation}
g_{ab}=g_{ab}^\text{(B)}+h_{ab}.
\end{equation}
The Sommerfeld boundary condition should then act like
\begin{equation}\label{eq:Sommerfeldh}
Lh_{ab}\doteq0.
\end{equation}
The use of approximations for $g_{ab}^\text{(B)}$ may introduce inaccuracies in the boundary condition, leading to junk reflections within the simulation domain.
To address this issue, we can further refine the boundary condition by taking a time derivative, thus removing the ambiguity in $g_{ab}^\text{(B)}$, resulting in
\begin{equation}
\partial_t\left(Lg_{ab}\right)\doteq0.
\end{equation}

In SpEC, the Einstein equations are solved using the generalized harmonic (GH) formalism, with the gauge condition given by
\begin{equation}\label{eq:GH}
\nabla^c\nabla_c x^a =H^a,
\end{equation}
where $H^a$ are the gauge source functions.
In the GH formalism, the Einstein equations can be written in first-order form \cite{Lindblom:2005qh} by introducing the derivatives of $g_{ab}$ as additional variables $\Phi_{iab}\equiv\partial_ig_{ab}$ and $\Pi_{ab}\equiv-t^c\partial_cg_{ab}$, where $t^c$ is the future directed unit normal to the $t=\text{constant}$ hypersurfaces.
The gauge boundary condition is imposed on the incoming characteristic field, given by
\begin{equation}
u_{ab}^{1-}=\Pi_{ab}-n^i\Phi_{iab}-\gamma_2g_{ab},
\end{equation}
where $n^i$ is the outward-pointing spatial normal vector to the outer boundary, and $\gamma_2$ is introduced to damp the constraint violation $\partial_ig_{ab}-\Phi_{iab}$.
The ingoing and outgoing null vectors are given by $k^a=\frac{1}{\sqrt{2}}\left(t^a-n^a\right)$ and $l^a=\frac{1}{\sqrt{2}}\left(t^a+n^a\right)$.
In this notation, if we assume the background metric to be either constant or changes ``adiabatically'', i.e., $\partial_t g_{ab}^\text{B} \sim \frac{v}{c} \partial_t h_{ab}$, where $v$ is the coordinate velocity of the BH and $v/c\ll1$, and $h_{ab}$ is a spherical wave, then the Sommerfeld gauge boundary condition reads approximately as
\begin{equation}\label{eq:OriginalBC}
P^{\text{G}cd}_{ab}d_t\left[u_{cd}^{1-}+\left(\gamma_2-\frac{1}{r}\right)g_{cd}\right]\doteq0.
\end{equation}
Here $P^{\text{G}cd}_{ab}=\left[k_ak_bl^{(c}+k_a\delta_b^{(c}+k_b\delta_a^{(c}\right]l^{d)}$ is the projection operator that removes the four constraint-preserving degrees of freedom corresponding to the incoming constraint fields and two physical degrees of freedom that correspond to backscattered gravitational waves. This leaves four degrees of freedom that correspond to the gauge part of the field, which determine the lapse and shift (see \cite{Rinne:2007ui} for more details).
Here $d_t u$ is the incoming characteristic projection of $\partial_t u$,
  as defined in Ref.~\cite{Lindblom:2005qh}:
  If the fundamental fields are $u^\alpha$ and the characteristic fields
  are $u^{\hat \alpha} \equiv e^{\hat \alpha}{}_{\beta} u^{\beta}$ where
  Greek letters label each field and
  $e^{\hat \alpha}{}_{\beta}$ are the characteristic eigenvectors,
  then $d_t u^{\hat \alpha} \equiv e^{\hat \alpha}{}_{\beta} \partial_t u^{\beta}$.

\subsection{Reflection from the boundary condition}
To understand how the CoM drift is coupled with the outer boundary condition, we first consider a scalar wave in 3-dimensional flat space as a toy model:
\begin{multline}
\psi_{\ell mk}(r,\theta,\phi,t)=e^{ikt}\Big(h^{(1)}_\ell(k|\bm{\mathrm{r}}-\bm{\mathrm{r_s}}|)Y^m_\ell(\theta,\phi)\\
+\sum_{\ell',m'}\rho^\ell_{\ell' m'}h^{(2)}_{\ell'}(k|\bm{\mathrm{r}}-\bm{\mathrm{r_s}}|)Y^{m'}_{\ell'}(\theta,\phi)\Big),
\end{multline}
where $h^{(1)}, h^{(2)}$ are spherical Hankel functions that correspond to the outgoing and ingoing waves, $\rho^\ell_{\ell' m'}$ are the reflection coefficients, and $\bm{\mathrm{r_s}}$ is the location of the source.

\begin{figure}[thb]
\centering
\includegraphics[width=0.45\textwidth]{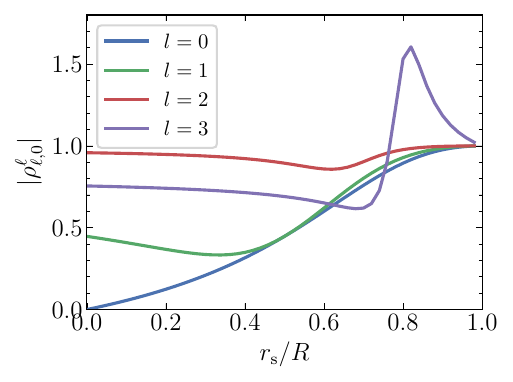}
\caption{\justifying Reflection coefficient $|\rho^\ell_{\ell' m'}|$ for different $\ell$. Given $m'=0, \ell'=\ell, k=1, t=0, \theta=\phi=0$.}
\label{fig:scalar1}
\end{figure}

We impose the Sommerfeld boundary condition Eq.~\eqref{eq:Sommerfeld} on this scalar field, and calculate $\rho^\ell_{\ell' m'}$ for $0\leq\ell\leq3$, $\ell'=\ell$ and $m'=0$.
In Fig.~\ref{fig:scalar1}, we show the results as a function of the distance of the source from the origin, $r_\text{s}$, normalized by the outer boundary radius $R$.
We can see that the Sommerfeld boundary condition is only perfectly non-reflective for the $\ell=0$ mode for a source located at the origin ($r_\text{s}=0$).
When the source is positioned away from the origin ($r_\text{s}>0$), there will be mode mixing and the Sommerfeld boundary condition is not non-reflective anymore.
As we shall show below in Sec.~\ref{sec:EoM}, this reflection of the gauge wave leads to the drift of the CoM observed in BBH simulations.

This leads us to generalize the Sommerfeld boundary condition in Eq.~\eqref{eq:Sommerfeld} by introducing a parameter $\bm{\mathrm{r_0}}$.
The radial coordinate $r$ is modified to $|\bm{\mathrm{r}}-\bm{\mathrm{r_0}}|$, in order to eliminate the reflections of $\ell\neq0$ modes
\begin{equation}\label{eq:scalarBC}
L(\bm{\mathrm{r_0}})\psi\doteq0,
\end{equation}
where
\begin{equation}\label{eq:SommerfeldOpGen}
L(\bm{\mathrm{r_0}})=\partial_t+\partial_{|\bm{\mathrm{r}}-\bm{\mathrm{r_0}}|}+\frac{1}{|\bm{\mathrm{r}}-\bm{\mathrm{r_0}}|}.
\end{equation}
$L(0)\psi\doteq0$ corresponds to the original Sommerfeld boundary condition.

\begin{figure}[thb]
\centering
\includegraphics[width=0.45\textwidth]{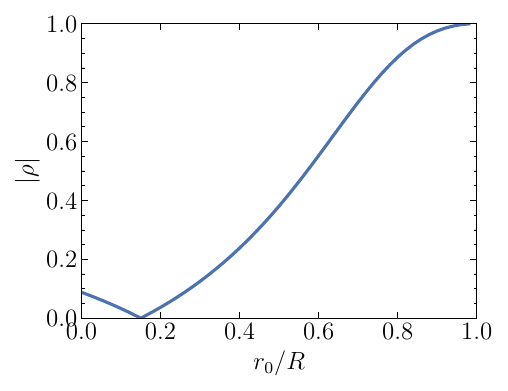}
\caption{\justifying Reflection coefficient $|\rho|$ for a spherical wave whose source
  is localized at $\bm{\mathrm{r_s}}/R=0.15$, with $k=1, t=0, \theta=\phi=0$.
  \label{fig:scalar2}
}
\end{figure}
To show how the introduced parameter $\bm{\mathrm{r_0}}$ works, we consider a spherical wave sourced at $\bm{\mathrm{r_s}}$ instead of the origin,
\begin{equation}
\psi(\bm{\mathrm{r}},t)=\frac{1}{|\bm{\mathrm{r}}-\bm{\mathrm{r_s}}|}\left(e^{ik(|\bm{\mathrm{r}}-\bm{\mathrm{r_s}}|-t)}+\rho e^{-ik(|\bm{\mathrm{r}}-\bm{\mathrm{r_s}}|+t)}\right),
\end{equation}
which has nonzero $\ell$ modes with respect to the origin.
In this case we can adjust the $\bm{\mathrm{r_0}}$ parameter in Eq.~\eqref{eq:SommerfeldOpGen} to shift the boundary condition with respect to the source, effectively eliminating reflected waves.
Fig.~\ref{fig:scalar2} illustrates the reflection coefficient $\rho$ as a function of $\bm{\mathrm{r_0}}$. 
Notably, when $\bm{\mathrm{r_0}}=\bm{\mathrm{r_s}}$, the boundary condition becomes non-reflective, demonstrating its efficacy in preventing wave reflections.

\section{CoM dynamics}\label{sec:EoM}
In this section we will derive the leading-order equation of motion for the CoM under the interaction with the outer boundary gauge reflection.
The derivation will only focus on the leading-order effects.

\begin{figure}[thb]
\centering
\includegraphics[width=0.45\textwidth]{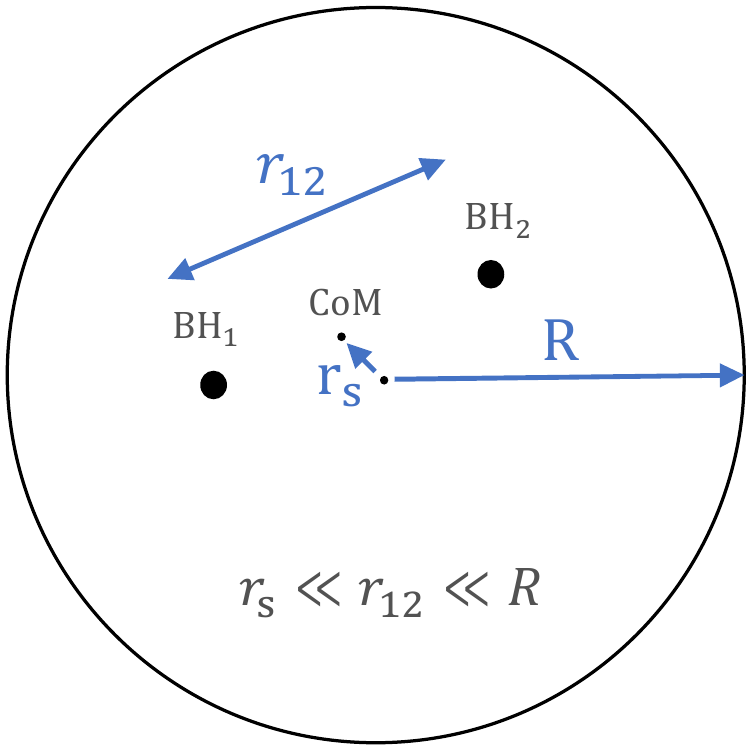}
\caption{\justifying A schematic diagram showing the domain setup and symbols.}
\label{fig:schematic}
\end{figure}

We use the following notation: $\bm{\mathrm{r_s}}$ is the location of the CoM.
$\bm{\mathrm{R}}$ is the field point on the outer boundary,
and $\bm{\mathrm{x_1}}$ and $\bm{\mathrm{x_2}}$ are
the locations of the black holes.
We assume $\mathrm{r_s} \ll r_{12}= |\bm{\mathrm{x_1}}-\bm{\mathrm{x_2}}| \ll R$, and $\mathrm{r_s} \ll 1/k$, where $k$ is the wavenumber of the metric perturbation.
The setup is shown in the schematic diagram in Fig.~\ref{fig:schematic}.
The metric perturbation in the region near the outer boundary ($r\sim R\gg r_{12}$) can be written as the summation of outgoing and ingoing waves
\begin{multline}
h_{ab}(\bm{\mathrm{r}},t)=h_{ab}^{\text{outgoing}}(\bm{\mathrm{r}},t)+h_{ab}^{\text{ingoing}}(\bm{\mathrm{r}},t)\\
{}=\Big(f^\text{out}_{ab}(\theta,\phi)\frac{e^{ik|\bm{\mathrm{r}}-\bm{\mathrm{r_s}}|}}{|\bm{\mathrm{r}}-\bm{\mathrm{r_s}}|}
+f^\text{in}_{ab}(\theta,\phi)\sum_{\ell'}\rho_{\ell'}h_{\ell'}^{(2)}\left(kr\right)\Big)e^{-ikt}\\
+O\left(\frac{1}{R^2}\right).
\end{multline}
Near the outer boundary, the quantity
$h_{\ell'}^{(2)}\left(kr\right)$ behaves like $e^{-ikr}/r^{\ell+1}$, so we will only focus on the dominant $\ell=0$ reflected mode.
Substituting the Sommerfeld boundary condition Eq.~\eqref{eq:Sommerfeldh}, we obtain
\begin{equation}
\rho_{0}=\frac{\hat{\bm{\mathrm{R}}}\cdot\bm{\mathrm{r_s}}}{2R^2}e^{2ikR}+O\left(\frac{r_s^2}{R^2}\right).
\end{equation}
The ingoing perturbation is then given by
\begin{equation}\label{eq:hingoingR}
h_{ab}^{\text{ingoing}}(\bm{\mathrm{r}},t)=f_{ab}^\text{in}(\theta,\phi)\frac{\hat{\bm{\mathrm{r}}}\cdot\bm{\mathrm{r_s}}}{2R^2}\frac{e^{ik(2R-r-t)}}{r}+O\left(\frac{r_s^2}{R^3}\right).
\end{equation}

The momentum flux carried by a gravitational perturbation on the outer boundary is given by the integral of the Landau–Lifshitz pseudotensor on the outer boundary
\begin{equation}\label{eq:MomentumFlux}
\frac{dP^i}{dt}=-\int_{r=R} t_\text{LL}^{ij}\hat{r}_jr^2d\Omega,
\end{equation}
where
\begin{widetext}
\begin{equation}\label{eq:LL}
\begin{aligned}
t_\text{LL}^{ij}=\frac{1}{16\pi(-g)}&\left[\pd_a\left(\sqrt{-g}g^{ij}\right)\pd_b\left(\sqrt{-g}g^{ab}\right)-\pd_a\left(\sqrt{-g}g^{ia}\right)\pd_b\left(\sqrt{-g}g^{jb}\right)\right.\\
&\left.+\frac{1}{2}g^{ij}g_{ab}\pd_c\left(\sqrt{-g}g^{ad}\right)\pd_d\left(\sqrt{-g}g^{bc}\right)+g_{ab}g^{cd}\pd_c\left(\sqrt{-g}g^{ia}\right)\pd_d\left(\sqrt{-g}g^{jb}\right)\right.\\
&\left.-\left(g^{ia}g_{bc}\pd_d\left(\sqrt{-g}g^{jc}\right)\pd_a\left(\sqrt{-g}g^{bd}\right)+g^{ja}g_{bc}\pd_d\left(\sqrt{-g}g^{ic}\right)\pd_a\left(\sqrt{-g}g^{bd}\right)\right)\right.\\
&\left.+\frac{1}{8}\left(2g^{ia}g^{jb}-g^{ij}g^{ab}\right)\left(2g_{cd}g_{ef}-g_{de}g_{cf}\right)\pd_a\left(\sqrt{-g}g^{cf}\right)\pd_b\left(\sqrt{-g}g^{de}\right)\right]
\end{aligned}
\end{equation}
\end{widetext}
is quadratic in the first derivatives of the metric.
Using the decomposition
\begin{equation}
g_{ab}=g^{\text{(B)}}_{ab}+h^\text{outgoing}_{ab}+h^\text{ingoing}_{ab},
\end{equation}
the terms in the outer boundary integral in Eq.~\eqref{eq:MomentumFlux} that involve only $g^\text{(B)}_{ab}$ and $h^\text{outgoing}_{ab}$ drop out, because they are even functions with respect to the source when time-averaged over orbits, while $\hat{r}$ is odd, therefore their product is odd and integrates to zero over the boundary.
So the leading-order non-vanishing result of the integral Eq.~\eqref{eq:MomentumFlux} comes from terms involving the product of $\partial_ag^\text{(B)}_{bc}$ and $\partial_ah^\text{ingoing}_{bc}$.
The terms involving the product of $\partial_ah^\text{outgoing}_{bc}$ and $\partial_ah^\text{ingoing}_{bc}$ are on the order of $\mathcal{O}(h^2)$, so we will ignore them in the leading-order calculation.

In the region near the outer boundary ($r\sim R\gg r_{12}$), the background metric can be described using the post-Newtonian (PN) approximation \cite{Blanchet1989PostNewtonianGO,Blanchet_1998}
\begin{equation}
\begin{aligned}
g^\text{(B)}_{00}=&-1+\frac{2Gm_1}{c^2r_1}+O\left(\frac{1}{c^4}\right)+1\leftrightarrow2,\\
g^\text{(B)}_{0i}=&-\frac{4m_1}{c^3r_1}v_1^i+O\left(\frac{1}{c^5}\right)+1\leftrightarrow2,\\
g^\text{(B)}_{ij}=&\left(1+\frac{2Gm_1}{c^2r_1}\right)\delta_{ij}+O\left(\frac{1}{c^4}\right)+1\leftrightarrow2.
\end{aligned}
\end{equation}
So we have $\pd_a g^\text{(B)}_{bc} \sim 1/R^2+O(v/R^2)$.
And from Eq.~\eqref{eq:hingoingR}, we have $\pd_a h^\text{ingoing}_{bc} \sim (\hat{\bm{\mathrm{R}}}\cdot r_\text{s})/R^3 + O(r_\text{s}^2/R^4)$.
Thus Eq.~\eqref{eq:MomentumFlux} gives $dP^i/dt \sim r_\text{s}^i/R^3$ at leading order.
This leads us to the CoM equation of motion:
\begin{equation}\label{eq:EoM0}
\ddot{\bm{\mathrm{r}}}_{\text{s}(t)}-\frac{a}{R^{3}} \bm{\mathrm{r}}_{\text{s}(t_\text{ret})}=0,
\end{equation}
where $a$ is a dimensionless coefficient that collects the numerical factors from the leading-order evaluation of the outer boundary Landau-Lifshitz momentum flux, after keeping only the dominant $\partial g^{\text{(B)}} \partial h^\text{ingoing}$ contribution.
Here, the momentum flux takes the value at a retarded time $t_\text{ret}\approx t-2R$.
For $t\gg R$, the solution is given by
\begin{equation}
\bm{\mathrm{r}}_{\text{s}}(t)=\bm{\mathrm{A}}\exp\left(\frac{\sqrt{a}}{R^{3/2}}t\right).
\end{equation}
The solution with a positive $a$ is consistent with \cite{Szil_gyi_2015}, whose results showed the exponential growth rate $\sigma\propto R^{-1.45}$, as shown in Fig.~\ref{fig:ExpFit}.
In this leading order model the binary parameters primarily set the magnitude of the flux rather than its parity structure, so we do not expect the sign of the coefficient $a$ to change with binary parameters.

We emphasize that the mechanism underlying Eq.~\eqref{eq:EoM0} requires the outer boundary to be causally connected with the source (i.e., that reflected gauge perturbations can return on a timescale $\sim 2R$).
If instead the outer boundary is placed sufficiently far out that the evolution time satisfies
$t_{\rm evol}\lesssim 2R$, the feedback loop is effectively absent.

\section{Modified boundary conditions}\label{sec:NewBC}
Having understood how the Sommerfeld boundary condition in Eq.~\eqref{eq:scalarBC} leads to exponential drift of the CoM, we now use Eq.~\eqref{eq:SommerfeldOpGen} to modify the boundary condition to control the gauge reflection and the CoM drift.
The parameter $\bm{\mathrm{r_0}}$ is introduced as a source term in the boundary condition.
Following the derivation in Sec.~\ref{sec:EoM}, we obtain the equation of motion for the CoM under Eq.~\eqref{eq:SommerfeldOpGen}
\begin{equation}\label{eq:CoMEoMRet}
\ddot{\bm{\mathrm{r}}}_{\text{s}(t)}+\frac{a}{R^{3}} \left(\bm{\mathrm{r}}_{\bm{0}}-\bm{\mathrm{r}}_{\bm{\text{s}}}\right)_{(t_\text{ret})}=0.
\end{equation}
For $t\gg R$, the second term can be linearly expanded as
\begin{equation}\label{eq:CoMEoM}
\ddot{\bm{\mathrm{r}}}_{\text{s}}-\frac{2b}{R^{2}} \left(\dot{\bm{\mathrm{r}}}_{\bm{0}}-\dot{\bm{\mathrm{r}}}_{\text{s}}\right)+\frac{a}{R^{3}} \left(\bm{\mathrm{r_0}}-\bm{\mathrm{r}}_{\text{s}}\right)=0,
\end{equation}
where $b=(t-t_\text{ret})/(2R)$.

We approximately treat $b$ as a constant parameter, given that the location of the outer boundary doesn't change significantly during the simulation, i.e., $\Delta R/R \ll1$. Then we can treat the CoM coordinate quantities (i.e., $\bm{\mathrm{r}}_\text{s}$, $\dot{\bm{\mathrm{r}}}_{\bm{\text{s}}}$ and $\ddot{\bm{\mathrm{r}}}_{\text{s}}$) as measured at the time when the boundary condition is imposed, rather than at the retarded time.

One possible approach to control the drift is to allow the auxiliary parameter $\bm{\mathrm{r_0}}$ above to depend on $\bm{\mathrm{r_\text{s}}}$, $\dot{\bm{\mathrm{r}}}_{\bm{\text{s}}}$ and $t$ in a way that ensures the solution to Eq.~\eqref{eq:CoMEoM} does not exhibit exponential growth.
In the following, we consider two specific forms of such dependence.

\subsubsection{Example 1}
We take a simple case as an example.
Consider $\bm{\mathrm{r_0}}=\lambda\bm{\mathrm{r}}_{\bm{\text{s}}}$, where $\lambda$ is a free parameter.
Substituting into the CoM equation of motion Eq.~\eqref{eq:CoMEoM}, we get
\begin{equation}\label{eq:eom1}
\ddot{\bm{\mathrm{r}}}_{\text{s}}-2 \sigma \dot{\bm{\mathrm{r}}}_{\bm{\text{s}}}+\omega^2 \bm{\mathrm{r}}_{\text{s}}=0,
\end{equation}
where
\begin{equation}\label{eq:SigmaOmega}
\sigma=b \frac{(\lambda-1)}{R^{2}},\quad
 \omega=\sqrt{a} \frac{(\lambda-1)^{1/ 2}}{R^{1.5}}.
\end{equation}
For the limit $R\gg r_\text{s}$ and $t\gg 1$, the equation gives
\begin{equation}\label{eq:sol1}
\bm{\mathrm{r}}_{\text{s}}=\left\{\begin{array}{ll}
\bm{\mathrm{A}} e^{|\omega|t} & (\lambda<1), \\
\bm{\mathrm{A}} e^{\sigma t}\sin\left(\omega t+\phi_0\right) & (\lambda>1).
\end{array}\right.
\end{equation}
In terms of the incoming characteristic field, the boundary condition Eq. ~\eqref{eq:SommerfeldOpGen} reads
\begin{equation}\label{eq:newBC1}
\begin{aligned}
P_{a b}^{\mathrm{G} c d} d_{t}&\left[ u_{c d}^{1-}+\left(n^{i}-\frac{r^{i}-\lambda r_\text{s}^{i}}{|\bm{\mathrm{r}}-\bm{\mathrm{r_s}}|}\right) \Phi_{i c d}\right. \\
&\quad+\left.\left(\gamma_{2}-\frac{1}{|\bm{\mathrm{r}}-\lambda\bm{\mathrm{r_s}}|}\right) g_{c d}\right] \doteq 0.
\end{aligned}
\end{equation}

\begin{figure}[thb]
\centering
\includegraphics[width=0.45\textwidth]{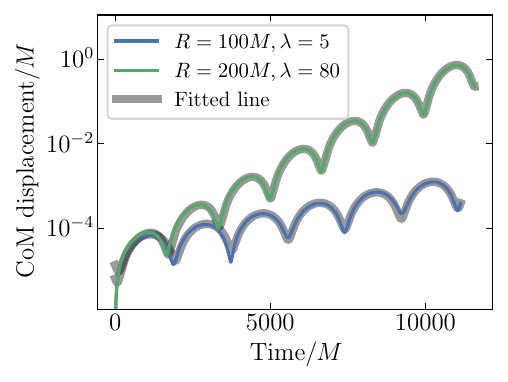}
\caption{\justifying The green and blue lines are the results for $R=100\,M, \lambda=5$ and  $R=200\,M, \lambda=80$, respectively, under gauge boundary condition Eq.~(\ref{eq:newBC1}). The wider gray lines are there fits. The system is equal-mass and non-spinning. The initial separation is $19\,M$ and the merger time is around $11000\,M$.}
\label{fig:fit}
\end{figure}

To test the model, we perform simulations of equal-mass and non-spinning system with different outer boundary radii $R$ and free parameter $\lambda$.
The initial separation is $19\,M$ and the merger time is around $11000\,M$.
The simulation results exhibit strong agreement with the model fitting based on Eq.~\eqref{eq:sol1}, as shown in Figure \ref{fig:fit}.
By fitting the growth rate $\sigma$ and frequency $\omega$ from Eq.~\eqref{eq:SigmaOmega} and \eqref{eq:sol1} to the NR results for different $\lambda$, we obtain the parameters $a=0.54$ and $b=0.64$, which are independent of $\lambda$.
The parameter $b$ is greater than $a$ due to the GR time lag.

\begin{figure}[thb]
\centering
\includegraphics[width=0.45\textwidth]{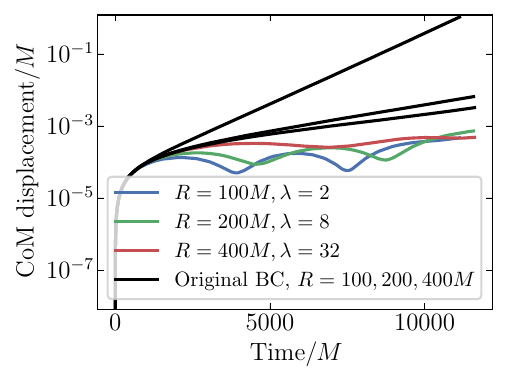}
\includegraphics[width=0.45\textwidth]{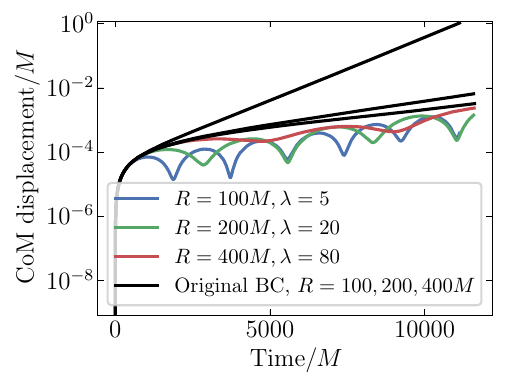}
\caption{\justifying Different results with similar $(\lambda-1)/R^{2}$ ratio under gauge boundary condition Eq.~(\ref{eq:newBC1}). The system is equal-mass and non-spinning. The initial separation is $19\,M$ and the merger time is around $11000\,M$.}
\label{fig:test}
\end{figure}

To further test this model, we consider the growth rate $\sigma=0.64 (\lambda-1)/R^{2}$.
According to this formulation, when $(\lambda-1)\sim R^{2}$, the growth rate is expected to be independent of $R$.
We plot the CoM drift for different $R$ and $\lambda$ in Fig.~\ref{fig:test}.
The observed results indeed agree with the anticipated behavior.

Eq.~\eqref{eq:sol1} shows that the growth rate of the solution is always positive, regardless of the value of the free parameter $\lambda$.
One way to decelerate the CoM is to introduce an additional term involving $\dot{\bm{\mathrm{r}}}_{\bm{\text{s}}}$ in the equation of motion, Eq.~\eqref{eq:eom1}, to act as a damping term.

To do this, we introduce an additional velocity term to Eq.~\eqref{eq:SommerfeldOpGen}:
\begin{equation}\label{eq:SommerfeldOpGenV}
L(\bm{\mathrm{r_0}},\dot{\bm{\mathrm{r}}}_{\bm{\text{s}}},\mu)=\partial_t+\partial_{|\bm{\mathrm{r}}-\bm{\mathrm{r_0}}|}+\frac{1}{|\bm{\mathrm{r}}-\bm{\mathrm{r_0}}|}+\mu\bm{\mathrm{n}}\cdot \dot{\bm{\mathrm{r}}}_{\bm{\text{s}}},
\end{equation}
where $\bm{\mathrm{n}}$ represents the outward-pointing normal vector to the outer boundary, and $\mu$ is a parameter.
With this modification, the boundary condition, expressed in terms of the incoming characteristic field, becomes:
\begin{equation}\label{eq:newBC}
\begin{aligned}
P_{a b}^{\mathrm{G} c d} d_{t}&\left[ u_{c d}^{1-}+\left(n^{i}-\frac{r^{i}-\lambda r_\text{s}^{i}}{|\bm{\mathrm{r}}-\lambda \bm{\mathrm{r_s}}|}\right) \Phi_{i c d}\right. \\
&\quad+\left.\left(\gamma_{2}-\frac{1}{|\bm{\mathrm{r}}-\lambda\bm{\mathrm{r_s}}|}\right) g_{c d}-\mu\frac{r^i-r_\mathrm{s}^i}{|\bm{\mathrm{r}}-\lambda\bm{\mathrm{r_s}}|}\dot{r}_{\mathrm{s}}^i\right] \doteq 0.
\end{aligned}
\end{equation}
Here $r_s$ and its time derivative are evaluated
  at the same coordinate time that the boundary condition is imposed.
The post-Newtonian equation of motion of the CoM under this new boundary condition is then
\begin{equation}
\left(1-2b\mu R\right)\ddot{\bm{\mathrm{r}}}_{\text{s}}+\left(\mu - 2 \sigma \right)\dot{\bm{\mathrm{r}}}_{\bm{\text{s}}}+\omega^2 \bm{\mathrm{r}}_{\text{s}}=0,
\end{equation}
where $b$ is the dimensionless retarded time parameter introduced in Eq.~\eqref{eq:CoMEoM}.
For this equation to exhibit damping behavior, the parameter $\mu$ must satisfy the following conditions:
\begin{equation}
\left\{\begin{array}{ll}
&\left(\mu-2\sigma\right)^2<4\omega^2(1-2b\mu R), \\
&2\sigma<\mu<\frac{1}{2bR}.
\end{array}\right.
\end{equation}
As a good choice, we can set
\begin{equation}
\mu\equiv\omega+2\sigma.
\end{equation}

\begin{figure}[thb]
\centering
\includegraphics[width=0.45\textwidth]{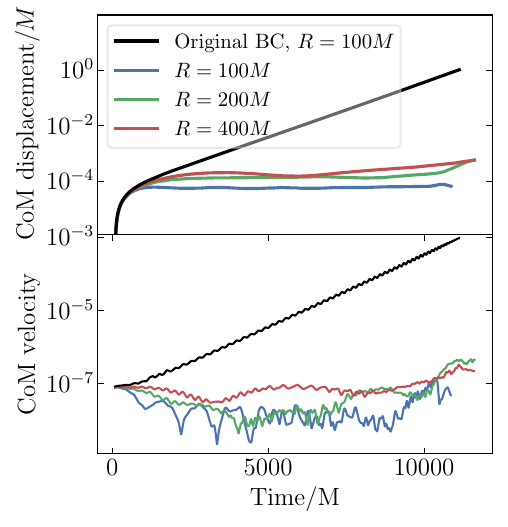}
\caption{\justifying CoM drift and velocity under the gauge boundary condition Eq.~\eqref{eq:newBC} with $\lambda=1+(R/100)^{2}$.
  The upper panel shows the CoM drift and the lower panel shows the CoM velocities. The black curves are the results obtained using the original Sommerfeld boundary condition, and the other curves are results with the new boundary condition for $R=100M, 200M$ and $400M$. The system is equal-mass and non-spinning. The initial separation is $19\,M$ and the merger time is around $11000\,M$.}
\label{fig:newBC}
\end{figure}
Now the only remaining free parameter is $\lambda$.
As discussed in Sec.~\ref{sec:EoM}, we can set $(\lambda-1)\sim R^{2}$ so that the growth or damping rate becomes independent of the outer boundary radius.
For $(\lambda-1)=(R/100)^{2}$, the corresponding results are displayed in Fig.~\ref{fig:newBC}.
This demonstrates the effectiveness of the modified boundary condition in damping the CoM drift.

\begin{figure}[thb]
\centering
\includegraphics[width=0.45\textwidth]{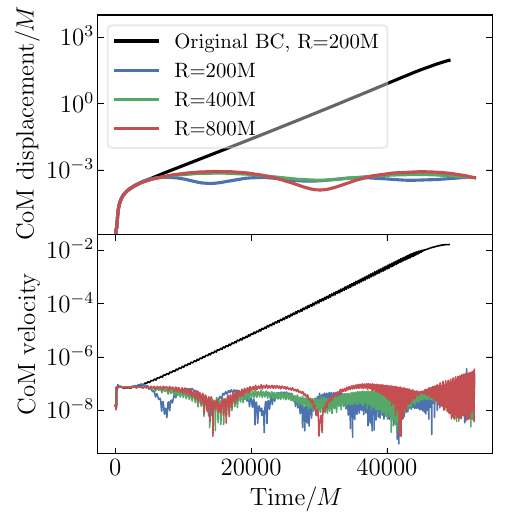}
\caption{\justifying CoM drift and velocity under gauge boundary condition Eq.~\eqref{eq:newBC1}
  with $\bm{\mathrm{r_0}}$ specified by Eq.~\eqref{eq:Varyr0}, where $\lambda=1+(R/100)^{2}$. The upper panel shows the CoM drift and the lower panel shows the CoM velocities. The black curves are the results obtained using the original Sommerfeld boundary condition, and other curves are results under the new boundary condition for $R=200M, 400M$ and $800M$. The system is equal-mass and non-spinning. The initial separation is $29\,M$ and the merger time is around $50000\,M$.}
\label{fig:eom1}
\end{figure}

An alternative way to introduce CoM damping for Eq.~\eqref{eq:eom1} without introducing an additional damping term like Eq.~\eqref{eq:SommerfeldOpGenV} is to use $\bm{\mathrm{r_0}}$ in the boundary condition only when the CoM is accelerating.
Namely,
\begin{equation}\label{eq:Varyr0}
\bm{\mathrm{r_0}}=\left\{\begin{array}{ll}
\lambda \bm{\mathrm{r}}_{\text{s}} & (\ddot{\bm{\mathrm{r}}}_{\text{s}} \cdot \dot{\bm{\mathrm{r}}}_{\text{s}}\geq 0), \\
0 & (\ddot{\bm{\mathrm{r}}}_{\text{s}} \cdot \dot{\bm{\mathrm{r}}}_{\text{s}}< 0).
\end{array}\right.
\end{equation}
The result for this approach is shown in Fig.~\ref{fig:eom1}, where we increased the initial separation to $29\,M$ and the merger time to around $50000\,M$ to show the long term behavior, since the oscillation time scale is longer than the previous cases.
This approach is more robust than Eq.~\eqref{eq:SommerfeldOpGenV} because it has fewer parameters.

\subsubsection{Example 2}
\begin{figure}[thb]
\centering
\includegraphics[width=0.45\textwidth]{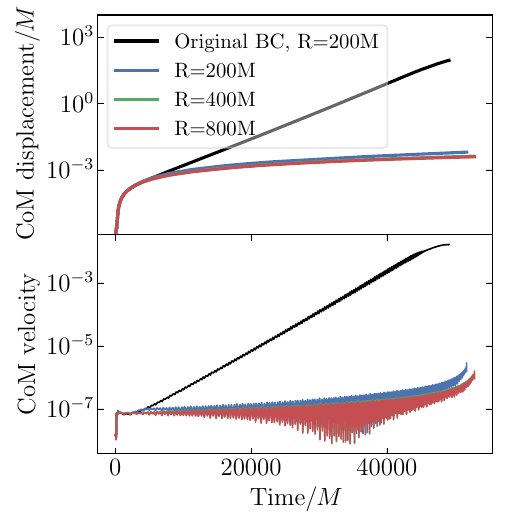}
\caption{\justifying CoM drift and velocity under gauge boundary condition Eq.~\eqref{eq:newBCExample2} with $\bm{\mathrm{r_0}}=\bm{\mathrm{r}}_{\bm{\text{s}}}+\eta R\dot{\bm{\mathrm{r}}}_{\bm{\text{s}}}$, where $\eta=R/4$. The upper panel shows the CoM drift and the lower panel shows the CoM velocities. The black curves are the results obtained using the original Sommerfeld boundary condition, and the other curves are results under the new boundary condition for $R=200M, 400M$ and $800M$. The system is equal-mass and non-spinning. The initial separation is $29\,M$ and the merger time is around $50000\,M$.}
\label{fig:eom2}
\end{figure}
Now we consider $\bm{\mathrm{r_0}}=\bm{\mathrm{r}}_{\bm{\text{s}}}+\eta R\dot{\bm{\mathrm{r}}}_{\bm{\text{s}}}$, where $\eta$ is a free parameter.
The boundary condition expressed in terms of the incoming characteristic field then reads
\begin{equation}\label{eq:newBCExample2}
\begin{aligned}
P_{a b}^{\mathrm{G} c d} d_{t}&\left[ u_{c d}^{1-}+\left(n^{i}-\frac{r^{i}-r_\text{s}^{i}-\eta R \dot{r}_{\mathrm{s}}^i}{|\bm{\mathrm{r}}-\bm{\mathrm{r_s}}-\eta R\dot{\bm{\mathrm{r}}}_{\bm{\text{s}}}|}\right) \Phi_{i c d}\right. \\
&\quad+\left.\left(\gamma_{2}-\frac{1}{|\bm{\mathrm{r}}-\bm{\mathrm{r_s}}-\eta R\dot{\bm{\mathrm{r}}}_{\bm{\text{s}}}|}\right) g_{c d}\right] \doteq 0.
\end{aligned}
\end{equation}
Then the CoM equation of motion Eq.~\eqref{eq:CoMEoM} becomes
\begin{equation}\label{eq:eom2}
\alpha\ddot{\bm{\mathrm{r}}}_{\text{s}}+\beta\dot{\bm{\mathrm{r}}}_{\bm{\text{s}}}=0,
\end{equation}
where
\begin{equation}
\alpha=1-\frac{2b\eta}{R},\quad \beta=\frac{a\eta}{R^2}.
\end{equation}
The equation gives
\begin{equation}\label{eq:sol2}
\bm{\mathrm{r}}_{\text{s}}=\bm{\mathrm{c_1}}-\bm{\mathrm{c_2}}e^{-\alpha t/\beta},
\end{equation}
where $\bm{\mathrm{c_1}}$ and $\bm{\mathrm{c_2}}$ are constants determined by the initial conditions.
Eq.~\eqref{eq:sol2} guarantees a damping solution as long as $\alpha>0$.
We plot the CoM drift and velocity under this gauge boundary condition in Fig.~\ref{fig:eom2}.
This approach avoids the oscillation of CoM around the coordinate origin, but the damping rate of the CoM drift is smaller than the other example.

\subsubsection{General system}
\begin{figure}[thb]
\centering
\includegraphics[width=0.45\textwidth]{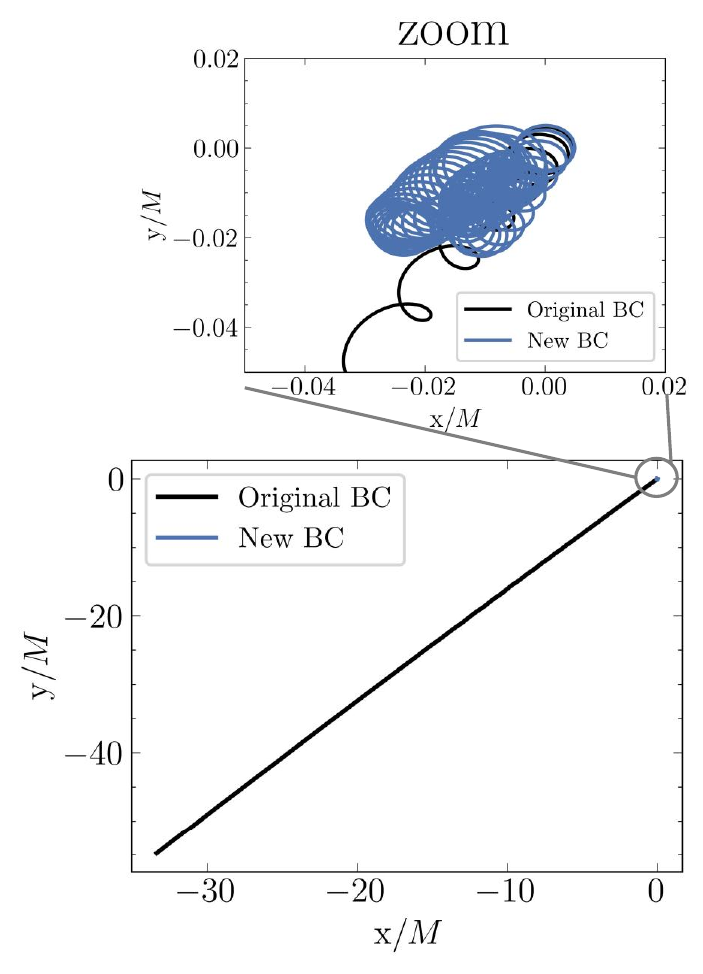}
\caption{\justifying CoM trajectory under gauge boundary condition Eq.~\eqref{eq:newBC1} with $\bm{\mathrm{r_0}}$ specified by Eq.~\eqref{eq:Varyr0}, where $\lambda=1+(R/100)^{2}$, with a zoom-in near the origin in the upper smaller panel. The black curve is obtained using the original Sommerfeld boundary condition, and the blue curve is obtained using the new boundary condition. The system has $q=1$, $\chi_1=(0.1,0.2,0.3)$, $\chi_2=(0.2,0.4,0.1)$, and $R=200\,M$. The initial separation is $29\,M$ and the merger time is around $65000\,M$. The vectorial displacement of the CoM under the new BC is shown in Fig.~\ref{fig:precessingxyz}, and the overall CoM drifting is shown in Fig.~\ref{fig:precessing}.}
\label{fig:circular}
\end{figure}

\begin{figure}[thb]
\centering
\includegraphics[width=0.45\textwidth]{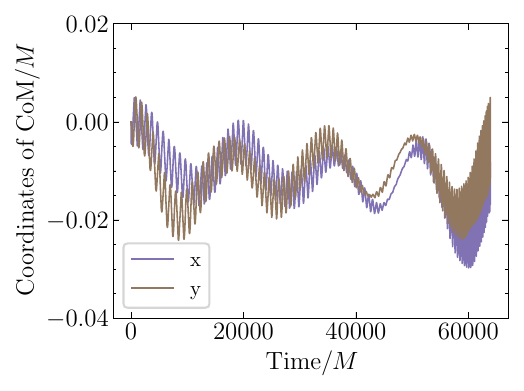}
\caption{\justifying The $x$ and $y$ coordinates of CoM under the new boundary condition with the same configuration with Fig.~\ref{fig:circular}.}
\label{fig:precessingxyz}
\end{figure}

\begin{figure}[thb]
\centering
\includegraphics[width=0.45\textwidth]{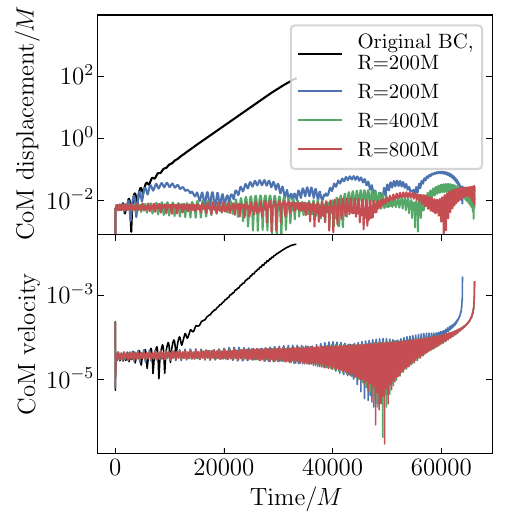}
\caption{\justifying CoM drift and velocity under gauge boundary condition Eq.~\eqref{eq:newBC1} with $\bm{\mathrm{r_0}}$ specified by Eq.~\eqref{eq:Varyr0}, where $\lambda=1+(R/100)^{2}$. The upper panel shows the CoM drift and the lower panel shows the CoM velocities. The black curves are the results obtained using the original Sommerfeld boundary condition, and the other curves are results under new boundary condition for $R=200M, 400M$ and $800M$. The system has $q=1$, $\chi_1=(0.1,0.2,0.3)$, and $\chi_2=(0.2,0.4,0.1)$. The initial separation is $29\,M$ and the merger time is around $65000\,M$. The black curves end earlier because the CoM displacement gets so large that the simulation crashes.}
\label{fig:precessing}
\end{figure}

The results obtained so far are specific to equal-mass non-spinning systems. 
In more general cases with non-equal masses and spin precession, the Newtonian measurement of CoM can exhibit circular motion in addition to the linear drift, as indicated by the blue curve in Fig.~\ref{fig:circular}.
However, in the new gauge boundary condition, Eq.~\eqref{eq:SommerfeldOpGen} or \eqref{eq:SommerfeldOpGenV}, we require the orbit-averaged drift coordinate and drift velocity as $\bm{\mathrm{r}}_{\text{s}}$ and $\dot{\bm{\mathrm{r}}}_{\bm{\text{s}}}$, rather than the quantities associated with the circular motion.
So we average the CoM coordinate and velocity over that last available orbit during the simulation, and take the averaged quantities $\langle\bm{\mathrm{r}}_{\text{s}}\rangle$ and $\langle\dot{\bm{\mathrm{r}}}_{\bm{\text{s}}}\rangle$ in the boundary condition.
Here $\langle f\rangle = \frac{\int^t_{t_0}f(t')dt'}{t-t_0}$, where $t_0$ is found by maximizing the dot product $\bm{\mathrm{r}}_{12}(t)\cdot\bm{\mathrm{r}}_{12}(t_0)$ over a time window $[t-3\pi/\Omega,t-\pi/\Omega]$, with $\Omega$ being the orbital angular velocity.

As an illustration, we take the boundary condition Eq.~\eqref{eq:SommerfeldOpGen} with Eq.~\eqref{eq:Varyr0}.
We use a system with mass ratio $q=1$ and dimensionless spin parameters $\chi_1=(0.1,0.2,0.3)$ and $\chi_2=(0.2,0.4,0.1)$. The initial separation is $29\,M$ and the merger time is around $65000\,M$.
The resulting trajectory is shown as the green curve in Fig.~\ref{fig:circular}, the vectorial displacement of the CoM under the new BC is shown in Fig.~\ref{fig:precessingxyz}, and the overall CoM drifting is shown in Fig.~\ref{fig:precessing}.
The surge of the CoM velocity near merger is expected and physical: it is the kick due to the asymmetry of the BBH system.
Testing shows that this approach is effective for mass ratio $q$ up to $8$ and spin $|\chi_{1,2}|$ up to $0.8$.

\section{The effect on waveforms}\label{sec:waveform}

\begin{figure*}[th]
\begin{subfigure}[b]{0.45\textwidth}
\centering
\includegraphics[width=\textwidth]{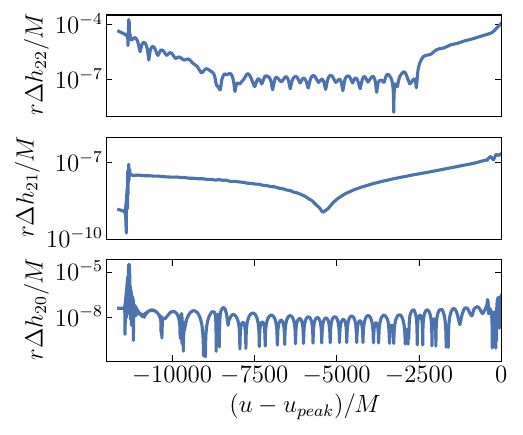}
\caption{\justifying The waveform difference for the equal-mass non-spinning case, $R=800\,M$. The matching window to fix the time shift and rotation starts from $8000\,M$ before merger, and is 10 orbits long. The BMS frame is fixed at the center of this matching window.}
\label{fig:waveform1}
\end{subfigure}
\hfill
\begin{subfigure}[b]{0.45\textwidth}
\centering
\includegraphics[width=\textwidth]{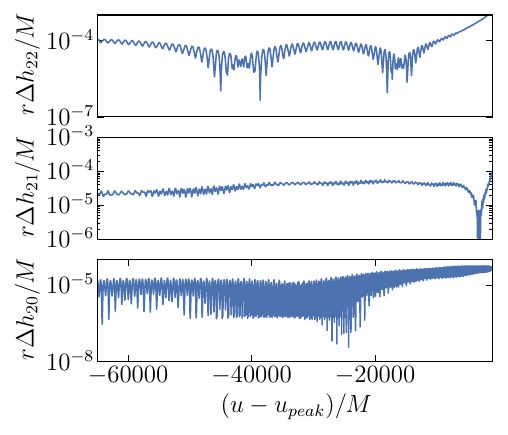}
\caption{\justifying The waveform difference for the precessing case with $q=1$, $\chi_1=(0.1,0.2,0.3)$, $\chi_2=(0.2,0.4,0.1)$, and $R=800\,M$. The matching window to fix the time shift and rotation starts from $58000\,M$ before merger, and is $50000\,M$ long. The BMS frame is fixed at the center of this matching window.}
\label{fig:waveform2}
\end{subfigure}
\caption{\justifying The waveform differences under modified gauge boundary condition Eq.~\eqref{eq:newBC1} with $\bm{\mathrm{r_0}}$ specified by Eq.~\eqref{eq:Varyr0}, where $\lambda=1+(R/100)^{2}$, and original boundary condition Eq.~\eqref{eq:OriginalBC} for two systems. We follow the procedure discussed in Sec. IV of \cite{Sun_2024} to fix the BMS frame, rotation and time shift between the two waveforms.
  \label{fig:waveform}
}
\end{figure*}

In theory,
modification to a gauge boundary condition should not have any
impact on waveforms. However, it is crucial to check this.
The conclusion should be general to all waveforms, not specific to a single example, since our boundary condition treatments do not depend on the intrinsic parameters of the system like masses and spins.

It is worth noting that the BBH system exhibits asymptotically flat behavior.  
Bondi, van~der~Burg, Metzner, and Sachs \cite{Sachs:1962wk,Sachs:1962zza,Bondi:1962px} demonstrated that the spacetime symmetry at the asymptotic boundary can be described by the Bondi-van~der~Burg-Metzner-Sachs (BMS) group.
This BMS group characterizes the asymptotic symmetry of spacetimes that are asymptotically flat at null infinity.
Refs.~\cite{Keefe:2021,Sun_2024} showed the importance of having both waveforms in the same BMS frame for making accurate comparisons.
We follow their process to map waveforms obtained using CCE to the same BMS frame using the \texttt{scri} package \cite{scri} and the \texttt{NRPNHybridization} code \cite{Sun:HybridizationWaveforms}.

After fixing the BMS frame and performing the required
3-D rotation and time alignment, we observe that the time-domain mismatch between the waveforms obtained under the Sommerfeld gauge boundary condition and our new gauge boundary condition for the equal-mass and non-spinning system over a $10000\,M$ window is on the order of $10^{-12}$, and for the precessing system over a $60000\,M$ window is on the order of $10^{-7}$. 
The residual is depicted in Fig.~\ref{fig:waveform} for these two cases. 
This mismatch is much smaller than the numerical error between different resolutions, since the numerical error is dominated by the grid spacing, and we are not changing the grids when modifying the boundary condition. The mismatch is consistent with the tolerance of BMS frame fixing and rotation alignment.
So there is no evidence suggesting that the modification of the gauge boundary condition can affect the waveform, once all the gauge ambiguities are fixed.
This is as expected since we only modify the gauge part of the boundary conditions.

\section{Alternative Boundary Conditions} \label{sec:alternative}
\subsubsection{Bayliss-Turkel condition}
Since Fig.~\ref{fig:scalar1} shows that the Sommerfeld boundary condition is only non-reflective for the $\ell=0$ mode, a straightforward idea is to use higher-order boundary conditions that are non-reflective for higher $\ell$ modes.
Such boundary conditions are known as Bayliss-Turkel boundary conditions \cite{Bayliss:1980}. They are given by
\begin{equation}
B_k\psi\doteq0,
\end{equation}
where
\begin{equation}\label{eq:BT}
B_k=\Pi^k_\ell\left(\partial_t+\partial_r+\frac{2\ell-1}{r}\right).
\end{equation}
However, given the presence of the high order derivatives, the well-posedness of these high-order boundary conditions when applied to the first order Generalized-Harmonic evolution \cite{Lindblom:2005qh} is not yet proven.

Here we test the second-order Bayliss-Turkel boundary condition. In terms of incoming characteristic fields in SpEC, it can be written as
\begin{multline}\label{eq:2ndBT}
P_{a b}^{\mathrm{G} c d}\left[\partial_t u_{c d}^{1-}-\partial_{t}\partial_{r} g_{c d}-\partial_{r}^2( g_{c d}-g^\text{background}_{c d})\right.\\
+\left(\gamma_{2}-\frac{4}{r}\right) \partial_{t} g_{c d}-\frac{4}{r} \partial_{r} ( g_{c d}-g^\text{background}_{c d})\\
\left.-\frac{2}{r^{2}} ( g_{c d}-g^\text{background}_{c d})\right]\doteq0.
\end{multline}
Here the first derivatives of the metric are computed using
  the fundamental variables $\Phi$ and $\Pi$ in the first-order
  GH system~\cite{Lindblom:2005qh},
  and the second derivatives of the metric are computed from derivatives
  of $\Phi$ and $\Pi$.
To approximate $g^\text{background}$, we tried using the Minkowski metric,
the Schwarzschild metric, and the metric used in the initial condition.
However, in all cases, there were large junk reflections that caused discontinuities in both CoM velocity and CoM drift, and even led to simulation crashes.
This raises concerns about the suitability of the second-order Bayliss-Turkel boundary conditions for BBH simulations in SpEC. 

An alternative approach to implement the Bayliss-Turkel type boundary condition without using the background metric is to introduce auxiliary variables defined on the outer boundary that evolve via a simple ODE system \cite{Buchman_2024}. These variables represent the residuals of the Bayliss–Turkel type absorbing operator Eq.~\eqref{eq:BT} for each $l,m$ mode at each order $k$. Driving them consistently to zero enforces the higher-order absorbing condition. In \cite{Buchman_2024}, the higher-order absorbing condition is only imposed on the physical degrees of freedom, and it is shown that the coordinate drift of CoM is not reduced.
We leave further investigation of imposing higher-order boundary conditions on the gauge degrees of freedom for future studies.

\subsubsection{Gauge constraint damping condition}
The gauge constraint damping boundary condition\cite{Lindblom:2009tu} explicitly drives the incoming characteristic fields on the outer boundary to satisfy the gauge condition $\mathcal{C}_a=H_a-\nabla^c\nabla_c x_a\rightarrow0$.

The gauge constraint damping boundary condition can be written as
\begin{equation}
d_t u^{1-}\doteq-\mu_B\left(u^{1-}-u^{1-}|_\text{BC}\right),
\end{equation}
where $\mu_B$ is a freely chosen damping parameter that exponentially suppresses any constraint violation crossing the boundary, and $u_{ab}^{1-}l^b|_\text{BC}$ is chosen so that the incoming characteristic projection of the constraint fields $\mathcal{C}_a$ vanishes.
In terms of the characteristic variables in SpEC,
the gauge constraint damping boundary condition reads

\begin{equation}\label{eq:GaugeConstraintDamping}
\begin{aligned}
P^{\text{G}cd}_{ab}d_tu_{cd}^{1-}&\doteq -\mu_BP^{\text{G}cd}_{ab}\left(u^{1-}_{cd}-u^{1-}_{cd}|_\text{BC}\right)\\
&\doteq\frac{\mu_B}{\sqrt{2}}\left(k_ak_bl^c+k_a\delta_b^c+k_b\delta_a^c\right)\left(\delta_c^d-t_ct^d\right)\mathcal{C}_d.
\end{aligned}
\end{equation}
Please refer to Appendix E of Ref.~\cite{Lindblom:2009tu} for details.

\begin{figure}[thb]
\centering
\includegraphics[width=0.45\textwidth]{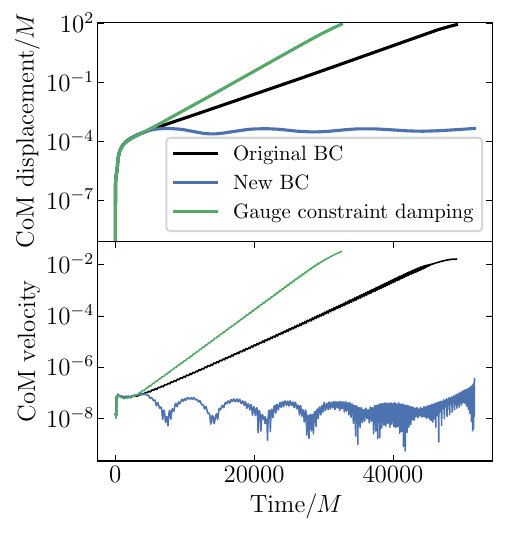}
\caption{\justifying CoM drift and velocity for an equal-mass and non-spinning system with an initial separation of $29\,M$, outer boundary radius $R=200\,M$, and the merger time is around $50000\,M$. The upper panel shows the CoM drift and the lower panel showx the CoM velocities. The black curves are the results obtained using the original Sommerfeld boundary condition Eq.~\eqref{eq:OriginalBC}. The blue curves are results under gauge boundary condition Eq.~\eqref{eq:newBC1}, with $\bm{\mathrm{r_0}}$ specified by Eq.~\eqref{eq:Varyr0}, where $\lambda=1+(R/100)^{2}$. The green curves are results under the gauge constraint damping boundary condition Eq.~\eqref{eq:GaugeConstraintDamping}. The green curves end earlier because the CoM displacement gets so large that the simulation crashes.}
\label{fig:GaugeConstraintDamping}
\end{figure}

The gauge constraint damping boundary condition should be
effective at suppressing violations of the GH gauge constraints at the outer boundary \cite{Lindblom:2009tu}.
Nevertheless, this condition does not enforce a purely outgoing wave: the gauge condition Eq.~\eqref{eq:GH} still admits both ingoing and outgoing gauge modes, so reflections can still occur in practice.
Fig.~\ref{fig:GaugeConstraintDamping} compares the CoM drift and its velocity for an equal‑mass, non‑spinning system with an initial separation of $29\,M$, and the merger time is around $50000\,M$.
Evolution that uses the gauge constraint damping boundary condition (blue curves) exhibits an even larger drift than the result obtained using the original Sommerfeld condition (black curves), indicating that the gauge constraint damping boundary condition does not reduce the reflection from the outer boundary and therefore fails to reduce the CoM drift.

\subsubsection{{Multipolar post-Minkowskian (MPM)}}
Another potential approach to reduce the CoM drift is to impose the boundary condition with analytic data supplied by the multipolar post-Minkowskian (MPM) expansion \cite{Thorne:1980ru,Blanchet1986,Blanchet_1998MPM,Blanchet_1995}.
Ref.~\cite{Blanchet_2021} gives the multipolar post-Minkowskian metric in Bondi coordinates, expressed entirely in terms of the source’s mass and current multipole moments.
By mapping the NR Cauchy coordinates to the Bondi frame using the CCE procedure developed in \cite{Moxon_2020}, one can evaluate the MPM solution on the outer boundary and supply it as time-dependent boundary data for the Cauchy evolution. 
Because this analytically matched metric already satisfies the correct outgoing-radiation behavior, it should suppress gauge reflections at the boundary and is therefore expected to reduce the CoM drift.
This procedure is significantly more complicated than other methods discussed here, so we leave it for future studies.

\subsubsection{{Cauchy-Characteristic matching (CCM)}}
Alternatively, Cauchy-Characteristic matching (CCM) \cite{dInverno,Bishop_1993,ma2024mergingblackholescauchycharacteristic,Ma_2024} can also provide accurate physical boundary conditions for the Cauchy domain by performing a matching of the internal Cauchy system and the exterior characteristic system.
But currently only the physical part of the fields is matched for BBH systems \cite{ma2024mergingblackholescauchycharacteristic,Ma_2024}.
Using CCM for the gauge part of the fields is a promising approach to reduce the CoM drift, and we leave this also for future studies.

\section{Conclusions}\label{sec:conclusions}
In this paper, we analyze the exponential drift of the CoM during long NR simulations of BBH systems. 
We have demonstrated that this drift is primarily caused by the reflection due to the gauge outer boundary conditions. 
To understand this issue, in the post-Newtonian approximation
we have derived the equation of motion for the CoM under various versions of the gauge boundary conditions.

To mitigate the drift, we propose a modification of the boundary conditions by introducing CoM source terms.
The specific form of the source term is not unique, and we present two specific examples showing that the CoM drift can be reduced by several orders of magnitude by incorporating these source terms into the gauge boundary condition.
\begin{acknowledgments}
We thank Nils Deppe, Kyle Nelli, Himanshu Chaudhary, Qing Dai and Harald Pfeiffer for useful discussions.
This work was supported in part by the Sherman Fairchild Foundation,
by NSF Grants
PHY-2309211, PHY-2309231, and OAC-2209656 at Caltech, and
PHY-2207342 and OAC-2209655 at Cornell.
Research
at Perimeter Institute is supported in part by the Government of
Canada through the Department of Innovation, Science and Economic
Development and by the Province of Ontario through the Ministry of
Colleges and Universities. Computations for this work were preformed on the Wheeler cluster at Caltech, which is supported by the Sherman Fairchild Foundation and by Caltech, and on the Resnick High Performance Computing (HPC) Cluster at the Caltech High Performance Computing Center.
\end{acknowledgments}

\def\bibsection{\section*{References}}
\bibliography{References}

\end{document}